\documentclass[a4paper]{spie}  

\usepackage{amsmath,amsfonts,amssymb}
\usepackage{graphicx}
\usepackage[colorlinks=true, allcolors=blue]{hyperref}
\usepackage{multirow}
\usepackage[utf8]{inputenc}
\usepackage{soulutf8}
\usepackage{subcaption}
\usepackage{booktabs}
\usepackage[separate-uncertainty=true,parse-numbers=true,binary-units=true]{siunitx}
\usepackage{tikz}
\usepackage{pgfplots}
\usepackage{pgfplotstable}
\usepackage{adjustbox}
\usepackage{bm}
\usepackage{nicefrac}
\pgfplotsset{compat=1.16}

\usepackage{cleveref} 

\usetikzlibrary{quotes,angles,calc}

\title{Recent advances in Bayesian optimization with applications to parameter reconstruction in optical nano-metrology}

\author[a]{Matthias Plock}
\author[a,b]{Sven Burger}
\author[a,b]{Philipp-Immanuel Schneider}
\affil[a]{Zuse Institute Berlin, Takustraße 7, 14195 Berlin, Germany}
\affil[b]{JCMwave GmbH, Bolivarallee 22, 14050 Berlin, Germany}
\renewcommand{\vec}{\mathbf}  

\begin{document}

\maketitle

\begin{abstract}
Parameter reconstruction is a common problem in optical nano metrology. It generally involves a set of measurements, to which one attempts to fit a numerical model of the measurement process. The model evaluation typically involves to solve Maxwell's equations and is thus time consuming. This makes the reconstruction computationally demanding. 
Several methods exist for fitting the model to the measurements. On the one hand, Bayesian optimization methods for expensive black-box optimization enable an efficient reconstruction by training a machine learning model of the squared sum of deviations $\chi^2$. On the other hand, curve fitting algorithms, such as the Levenberg-Marquardt method, take the deviations between all model outputs and corresponding measurement values into account which enables a fast local convergence. In this paper we present a Bayesian Target Vector Optimization scheme which combines these two approaches. We compare the performance of the presented method against a standard Levenberg-Marquardt-like algorithm, a conventional Bayesian optimization scheme, and the L-BFGS-B and Nelder-Mead simplex algorithms. As a stand-in for problems from nano metrology, we employ a non-linear least-square problem from the NIST Standard Reference Database. We find that the presented method generally uses fewer calls of the model function than any of the competing schemes to achieve similar reconstruction performance.
\end{abstract}


{\it \noindent
This paper will be published in Proc.~SPIE Vol.~{\bf 11783}
(2021) 117830J ({\it Modeling Aspects in Optical Metrology VIII}, DOI: 10.1117/12.2592266)
and is made available 
as an electronic preprint with permission of SPIE. 
One print or electronic copy may be made for personal use only. 
Systematic or multiple reproduction, distribution to multiple 
locations via electronic or other means, duplication of any 
material in this paper for a fee or for commercial purposes, 
or modification of the content of the paper are prohibited.}

\section{Introduction}
\label{sec:intro}

A parameter reconstruction in optical nano metrology is usually formulated as an inverse problem. After having recorded experimental measurement data, a parametric numerical model of the measurement process is created. A numerical optimization scheme seeks to minimize the deviation between the experimental and model results by repetitively evaluating the model for different parameter~\cite{Gross2006meas,Silver2008spie,Pang2012aot,Kumar2014oe}.

Depending on the numerical complexity of the model, obtaining the result for any given parameter can take from a few seconds to several hours. This creates the need for efficient and robust reconstruction algorithms that use the available resources most effectively, while still achieving excellent reconstruction results.

Bayesian optimization (BO) methods~\cite{movckus1975bayesian,mockus2012bayesian} have gained a reputation for being efficient at optimizing outputs of expensive models. They construct a stochastic surrogate of the model, mostly a Gaussian process (GP)~\cite{ki2006gaussian}, using the output of all previous model evaluations. At each iteration, the surrogate is used to identify new trial model parameters that maximize for example the expected improvement or the probability of improvement. The approach offers a large degree of flexibility such that many extensions of BO have been proposed to cope, e.g., with a large number of model parameters~\cite{wang2013bayesian,titsias2010bayesian}, a large number of observations~\cite{sant:2021}, or noisy model outputs~\cite{letham2019constrained}. 

Typically, BO methods are used to optimize scalar functions. In the case of a parameter reconstruction, often the sum of squared deviations between model and measurements $\chi^2$ is minimized. Minimizing the scalar $\chi^2$ has the drawback that the information on the multitude of model outputs is lost. On the other hand, least-square curve fitting methods, such as the Levenberg-Marquardt method, use all value outputs to fit the model to the measurement which allows for a fast local convergence. Recently, Uhrenholt \& Jensen proposed a method to perform such a nonlinear least-square regression using a BO method~\cite{uhrenholt2019efficient}.

In this paper, we present the application of this Bayesian target vector optimization (BTVO) scheme to a parameter reconstruction problem of the NIST Statistical Reference Database\cite{NIST_StRD}. This problem poses as a stand-in for actual problems from nano metrology, as it has a comparable number of parameters and degrees of freedom. While Uhrenholt \& Jensen compared their approach to the standard BO method, we extend this analysis in two aspects. Firstly, we consider the case that derivative information of the model is available to improve the efficiency of the parameter reconstruction. 
Secondly, we compare the performance to other standard algorithms used in metrology such as a Levenberg-Marquardt based least-square minimization, the Nelder-Mead algorithm and the gradient based L-BFGS-B method.

In \cref{sec:bayesian_opt_methods} we discuss BO methods. We first review a conventional BO scheme and discuss how it can be used to perform a least square regression task. We highlight the issues that can arise in this context. To solve these issues, we introduce the BTVO scheme. In \cref{sec:benchmark} we discuss the benchmarking procedure  to compare the proposed BTVO scheme to other common optimization methods. We introduce the problem for which we reconstruct the model parameters and show the reconstruction results for the different optimization methods. Here, we also show how the availability of derivative information can drastically speed up the parameter reconstruction. Finally, in \cref{sec:outlook}, we give a brief overview over possible applications and extensions which are made possible by the new BTVO method.

\section{Bayesian Optimization methods}
\label{sec:bayesian_opt_methods}

Bayesian optimization methods are global optimization methods that work very well for expensive black-box objective functions. In essence they consist of two components: the construction and training of a surrogate model at each step of the optimization, and the maximization of an acquisition function (which in turn uses the surrogate model) that then yields a new parameter $\vec{p}_{j+1}$ to sample. The goal of this iterative optimization process is to find a point $\vec{p}^{\ast}$ in the parameter space which maximizes or minimizes the objective function.

\subsection{Conventional Bayesian Optimization}
\label{sec:bo}
The conventional BO method is mainly used to optimize scalar functions. At the beginning of the optimization process, a stochastic surrogate model of the objective function $f: \mathcal{X}\subset\mathbb{R}^M \rightarrow \mathbb{R}$ is created and then trained with each new observation of $f(\vec{p})$. The surrogate model can be typically evaluated much faster than the expensive objective function. A common choice is to use GPs as surrogate models\cite{ki2006gaussian}. A GP is defined by a mean function $m(\vec{p})$ and a covariance kernel function $k(\vec{p}, \vec{p}^{\prime})$. Common initial choices, which are also employed in the following, are a constant mean function and a Mat\'ern 5/2 kernel function~\cite{brochu2010tutorial},
\begin{equation*}
    m(\vec{p}) = m_0\,, \quad \mathrm{and} \quad k(\vec{p},\vec{p}'; \boldsymbol{\theta}) = \sigma_0^2\left(1+\sqrt{5}r+\frac{5}{3}r^2\right)\exp\left(-\sqrt{5}r\right) \quad \mathrm{with} \quad r = \left( \sum_{i=1}^M \frac{(p_i-p_i')^2}{l_i^2} \right)^{\nicefrac{1}{2}} \,,
\end{equation*}
where $m_0$ and $[\sigma_{0}, l_{1}, \dots, l_{M}] = \boldsymbol{\theta}$ are the hyperparameters of the mean and covariance function, respectively. Given a set of observations $\vec{Y} = [\vec{y}_1, \vec{y}_{2}, \dots]$ for a set of model parameters $\vec{P} = [\vec{p}_1, \vec{p}_{2}, \dots]$ and hyperparameters $m_0,\boldsymbol{\theta}$, a GP can predict a normal distribution of the function value $\hat f(\vec{p}) \sim \mathcal{N}\left(\mu(\vec{p}),\sigma^{2}(\vec{p})\right)$ for each parameter $\vec{p}\in\mathcal{X}$ with
\begin{gather*}
    \mu(\vec{p}) = m_{0} + \vec{k}^{T} K^{(-1)} (\vec{Y} - \boldsymbol{1}m_{0}) \quad \mathrm{and} \quad \sigma^{2}(\vec{p}) = \sigma^{2}_{0} + \vec{k}^{T} K^{(-1)} \vec{k} \,, \\
    \mathrm{where} \quad \vec{k} = [k(\vec{p}, \vec{p}_{1}; \boldsymbol{\theta}), k(\vec{p}, \vec{p}_{2}; \boldsymbol{\theta}), \dots]^{T} \quad \mathrm{and} \quad (K)_{ij} = k(\vec{p}_{i}, \vec{p}_{j}; \boldsymbol{\theta}) \,.
\end{gather*}
The choice of the hyperparameter values is essential for obtaining a good prediction.  A common approach to optimize their valuesis to maximize the log marginal likelihood,
\begin{equation*}
    \log p(\vec{Y} | \vec{P}, \boldsymbol{\theta}) = -\frac{1}{2} \left(\vec{Y} - \boldsymbol{1}m_{0}\right)^{T} K^{(-1)}(\vec{P}; \boldsymbol{\theta}) \left(\vec{Y} - \boldsymbol{1}m_{0}\right) - \frac{1}{2}\log | K(\vec{P}; \boldsymbol{\theta}) | - \frac{n}{2}\log 2 \pi \,,
\end{equation*}
of the observed data $\vec{Y}$~\cite{ki2006gaussian,brochu2010tutorial}.
The predicted normal distribution is used to identify a new trial parameter set $\mathbf{p}_{j+1}$ for the next iteration of the optimization. To this end, an acquisition function $\alpha(\mathbf{p})$ is maximized, i.e.
\begin{equation}
    \mathbf{p}_{j+1} = \underset{\mathbf{p} \in \mathcal{X}}{\text{arg\,max}} \, \alpha(\mathbf{p}) \,.
\end{equation}
Common choices of acquisition functions are probability of improvement (PoI)~\cite{brochu2010tutorial}, expected improvement (EI)~\cite{movckus1975bayesian,jones1998efficient}, and lower- and upper confidence bound (LCB and UCB)~\cite{auer2002using}.

In order to perform a least-square regression with the conventional BO method, we have to scalarize the objective function. This is often done by calculating the reduced $\chi^2$ statistic,
\begin{equation}
    \label{eq:scalarized_chi_square}
    \chi_{\vec{p}}^2 = \frac{\left( \vec{y}_{\ast} - \vec{f}(\vec{p}) \right)^{T} H \left( \vec{y}_{\ast} - \vec{f}(\vec{p}) \right)}{N - M} \quad \text{with} \quad H = \mathrm{diag} \left(\boldsymbol{\eta}_{\ast}^{2} \right)^{-1} \,.
\end{equation}
Here, $\vec{y}_{\ast}$ are the observed data points to which we want to fit the model $\vec{f}(\vec{p})$, and $\boldsymbol{\eta}_{\ast}^2$ is a vector containing the measurement variance of the observed data points. Typically, the evaluation of the model function $\vec{f}(\vec{p})$ is computationally expensive, such that it can be beneficial to model $\chi_{\vec{p}}^2$ by a GP to determine parameter vectors $\vec{p}_{j+1}$ that likely lead to a reduction of $\chi_{\vec{p}}^2$.
However this approach has two important disadvantages. First, by modeling $\chi_{\vec{p}}^2$ with a single GP, most of the information about the $N$ dimensional model output of $\vec{f}(\vec{p})$ is lost. For example, the reduced $\chi^2$ statistic does not change when all differences between observed and modeled data points switch their respective sign. This information is potentially important for the optimization process and the prediction of a next sampling point. And second, GPs are designed to model data that follows a normal distribution. This is not the case for the reduced $\chi^2$ statistic, which is always larger or equal to zero. Both of these issues are addressed with the proposed BTVO method.

To mitigate the second problem, one can also consider bijective transformations $\chi_{\vec{p}}^2 \rightarrow g(\chi_{\vec{p}}^2)$ that change the distribution of function values. For large $N$, the transformation $g(\chi_{\vec{p}}^2) = \left(\chi_{\vec{p}}^2\right)^{\nicefrac{1}{3}}$ leads to approximately normally distributed function values~\cite{hawkins1986note}.

\subsection{Bayesian Target Vector Optimization}
\label{sec:blso}

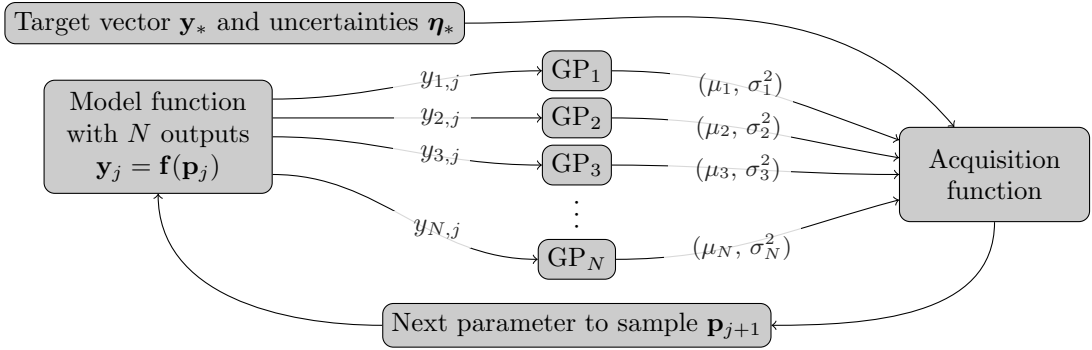
\begin{figure}[]
    \centering



\begin{tikzpicture}[]

  \node[draw, rectangle, minimum height=1.5cm, minimum width=3cm, fill=black!20,
  rounded corners, align=center] at (-0, 0) (sim) {Model function\\
    with $N$ outputs\\$\vec{y}_{j} = \vec{f}(\vec{p}_{j})$};

  \foreach \i in {1,...,3}{%
    
    \node[draw, fill=black!20, rounded corners, align=center] at
    ([xshift=4cm, yshift=1.5cm - \i * 0.625 cm]sim.east) (gp\i)
    {$\mathrm{GP}_{\i}$};
    
    \path[->] ([yshift=0.75cm - \i * 0.25 cm]sim.east) edge[out=0,in=180]
    node[pos=.6, fill=white, opacity=0.85, text opacity=1] {$y_{\i,j}$} (gp\i);

    ;}

  \node[draw, fill=black!20, rounded corners, align=center] at
  ([xshift=4cm, yshift=-1.625 cm]sim.east) (gpN)
  {$\mathrm{GP}_{N}$};
  
  \path[->] ([yshift=-0.5 cm]sim.east) edge[out=0,in=180] node[pos=.6, fill=white, opacity=0.85, text opacity=1]
  {$y_{N,j}$} (gpN);

  \node[rotate=90] at ($(gp3)!0.5!(gpN)$) (dots) {\dots};
  
  \node[draw, fill=black!20, rounded corners, align=center, minimum height=1.25cm, minimum width=2.5cm] at (11, -0.5)
  (af) {Acquisition\\ function};

  \node[draw, fill=black!20, rounded corners, align=center] at
  ([xshift=-0.5cm,yshift=0.75cm]sim.north east) (results) {Target vector
    $\vec{y}_{\ast}$ and uncertainties $\boldsymbol{\eta}_{\ast}$};

  \path[->] (results.east) edge[out=0,in=130] (af);

  \foreach \i in {1,...,3}{%
    
    \path[->] (gp\i.east) edge[out=0,in=150 + \i * 10] node[font=\small, pos=.4, fill=white, opacity=0.85, text opacity=1] {($\mu_{\i}$, $\sigma^2_{\i}$)} (af);

    ;}

  \path[->] (gpN.east) edge[out=0,in=195] node[font=\small, pos=.4, fill=white, opacity=0.85, text opacity=1] {($\mu_{N}$, $\sigma^2_{N}$)} (af);

  \node[draw, fill=black!20, rounded corners, align=center] at ([xshift=4cm, yshift=-2.5cm]sim.east)
  (cand) {Next parameter to sample $\vec{p}_{j+1}$};

  \path[->] (af) edge[out=270, in=0] (cand);

  \path[->] (cand.west) edge[out=180,in=270] (sim.south);
  
\end{tikzpicture}
    \caption{A schematic of the BTVO scheme. Each of the $N$ outputs of the model function is used to construct a surrogate model, in this case Gaussian processes. Each of these $N$ Gaussian processes provides predictions about the mean and variance in the parameter space, which is used by the acquisition function to calculate a new sampling point or evaluation candidate $\vec{p}_{j+1}$ for the model function.}
    \label{fig:tvo_schmatic}
\end{figure}

To overcome the issues found for a conventional BO approach, Uhrenholt \& Jensen propose to use $N$ individual GPs to model the $N$ outputs of the model function $\vec{f}(\vec{p})$~\cite{uhrenholt2019efficient} (cf. \cref{fig:tvo_schmatic} for a schematic overview), i.e.
\begin{equation*}
    \vec{y}_{j} = \vec{f}(\vec{p}_{j}) \,, \quad \mathrm{and} \quad y_{1,j} \to \mathrm{GP}_1 \,, \quad y_{2,j} \to \mathrm{GP}_2 \,, \quad \dots \,, \quad y_{N,j} \to \mathrm{GP}_N \,.
\end{equation*}
As GPs are a stochastic model of the objective function, each GP defines a normally distributed random variable $\hat f_i(\vec{p})\sim \mathcal{N}(\mu_{i}(\vec{p}),\sigma^{2}_{i}(\vec{p}))$ for each parameter $\vec{p}$ in the design space. Using these random variables, one can define a random variable 
\begin{equation*}
    \hat d_{i}(\vec{p}) = \frac{\hat f_{i}(\vec{p}) - y_{\ast, i}}{\eta_{\ast, i}} \,,
\end{equation*}
 for each difference from the model prediction to the respective target vector component $y_{\ast, i}$. Being a linear transformation of the GPs predictions, all $\hat d_{i}$ follow a normal distribution, with mean $d_i(\vec{p}) = (\mu_{i}(\vec{p}) - y_{\ast, i})/\eta_{\ast, i}$ and variance $\sigma_{i}^{2}(\vec{p}) / \eta^{2}_{\ast, i}$. The predicted residual sum of squares,
\begin{equation*}
    \hat{\chi}^2(\vec{p}) = \sum_{i}^{N} \hat{d}_{i}^{2}(\vec{p}) \,,
\end{equation*}
then follows the generalized $\chi^2$ distribution $G\chi^{2}(\vec{w}, \vec{N}, \boldsymbol{\lambda})$, with weight vector $\vec{w} = [\sigma_{1}^{2}(\vec{p}) / \eta_{\ast, 1}^{2}, \dots, \sigma_{N}^{2}(\vec{p}) / \eta_{\ast, N}^{2}]^T$, the degree-of-freedom vector $\vec{N} = [1, \dots, 1]^T$, and the non-centrality vector $\boldsymbol{\lambda} = [\lambda_1(\vec{p}), \dots, \lambda_{N}(\vec{p})]^T$, with each vector of length $N$. Here, the non-centrality parameter is defined as $\lambda_{i}(\vec{p}) = d_{i}^2(\vec{p})$. Predictions for the residual sum of squares require evaluation of the cumulative distribution function (CDF) of this probability distribution. Unfortunately, no closed form expression exists for the generalized $\chi^2$ distribution. Instead Uhrenholt \& Jensen consider the random variable
\begin{equation*}
    \tilde{\chi}^2(\vec{p}) = \sum_{i}^{N} \tilde{d}_{i}^{2}(\vec{p})\quad \text{with} \quad \tilde{d}_{i}^{2}(\vec{p}) = \hat{d}_{i}^{2}(\vec{p}) \frac{\eta_{\ast, i}^2}{\sigma_{i}^{2}(\vec{p})}\,.
\end{equation*}
Each rescaled squared distance $\tilde{d}_{i}^{2}(\vec{p})$ follows a non-central $\chi^2$ distribution $\tilde{d}_{i}^2(\vec{p}) \sim  NC\chi^2(1, \lambda(\vec{p}))$ with one degree of freedom and non-centrality parameter $\lambda(\vec{p}) = d_{i}^2(\vec{p}) \frac{\eta_{\ast, i}^2}{\sigma_{i}^{2}(\vec{p})}$.
Consequently, $\tilde{\chi}^2(\vec{p})$ follows a non-central $\chi^2$ distribution with $N$ degrees of freedom and non-centrality parameter 
\begin{equation*}
    \lambda(\vec{p}) = \gamma^{-2}(\vec{p}) \sum_{i}^{N} \left( \frac{\mu_{i}(\vec{p}) - y_{\ast, i}}{\eta_{\ast, i}} \right)^2 \quad \text{with} \quad
    \gamma^2(\vec{p}) = \frac{1}{N} \sum_{i}^{N} \frac{\sigma_{i}^{2}(\vec{p})}{\eta_{\ast, i}^{2}}\,.
\end{equation*}
The distribution $\gamma^2(\vec{p}) \tilde{\chi}^2(\vec{p})$ is then an unbiased approximation of the generalized non-central $\chi^2$ distribution of the random variable $\hat{\chi}^2(\vec{p})$~\cite{uhrenholt2019efficient}. It has the advantage that its CDF, denoted $F_{N, \lambda}(t)$, is known analytically. In order to compute acquisition functions such as the expected improvement and lower confidence bound efficiently, $F_{N, \lambda}(t)$ is replaced by its very accurate approximation by a Gaussian CDF,
\begin{equation*}
    F_{N, \lambda}(t) \approx \Phi \left( \frac{z - \alpha}{\varrho} \right) \,,
\end{equation*}
where $z = z(t, N, \lambda)$ 
refers to the random variable of the distribution, $\alpha = \alpha(t, N, \lambda)$ is the mean of the Gaussian CDF and $\varrho = \varrho(t, N, \lambda)$ is its standard deviation. With these approximations, the evaluations of the acquisition function require a comparable effort as for the conventional BO scheme. However, for large $N$ the computational effort to make predictions from $N$ independent GPs leads to a very large computational effort. To circumvent this issue we propose to use a shared covariance matrix $K$ across all GPs. The details of this approach will be further discussed in a future publication, which is currently in preparation.

\section{Benchmark}
\label{sec:benchmark}

In order to show the capabilities of the BTVO, we apply it to the problem of finding the optimal fit parameters in an analytical nonlinear least squares problem. We compare the proposed method against a Levenberg-Marquardt (LM) like scheme~\cite{levenberg1944method,marquardt1963algorithm} (we employ the scipy implementation \verb|scipy.optimize.least_squares|~\cite{2020SciPy-NMeth}), a conventional Bayesian optimization (BO) method (as described in \cref{sec:bo}), the L-BFGS-B scheme \cite{byrd1995limited} and the Nelder-Mead simplex algorithm (NM) \cite{nelder1965simplex}.

The BTVO and LM methods are both capable of natively performing a least squares regression, as they utilize the discrete outputs of the objective function to perform the fit. The BO, L-BFGS-B, and NM methods are usually employed to perform standard optimization tasks, where the minimum or maximum of a function is sought. In order to perform a least squares regression with these schemes, the objective function has to be scalarized first, as is described in \cref{sec:bo}.

Different optimization methods profit from different scalarization strategies. The L-BFGS-B scheme works very well on quadratic functions~\cite{liu1989limited}, while the BO implicitly assumes that the objective function is a random number that follows a normal distribution. By taking the third root of the $\chi^{2}$ distributed value \cref{eq:scalarized_chi_square}, we can achieve an approximately normally distributed variable~\cite{hawkins1986note}. Therefore when using the BO method we instead optimize $\left(\chi_{\vec{p}}^2\right)^{\nicefrac{1}{3}}$.

A benchmark consisting of six repeated runs was performed for each optimization method. From these results a mean and standard deviation is calculated, which is shown in the figures as a solid line and a shaded surrounding band of identical color, respectively. 
The reconstruction algorithms, which are implemented in the analysis and optimization toolbox of the finite element sofware JCMsuite\cite{jcmsuite}, were run on \num{8} cores of a Intel Xeon Gold 6246 CPU cores and \SI{8}{\giga\byte} of RAM per core.

Bayesian optimization methods can have a substantial computational overhead over other optimization methods. This is because to predict a new sample candidate, several smaller optimization problems have to be solved. This computational overhead may result in BO or BTVO requiring fewer iterations to reconstruct the optimal parameter, but still requiring a longer total runtime. Therefore both, a comparison of the number of iterations (where each iteration is one call to the objective function) and the total computation time are shown.

The metric we choose to measure the performance of the various optimization schemes is the distance $d$ of the reconstruction parameter $\vec{p}$ to the optimal reconstruction parameter $\vec{p}^{\ast}$. The distance in each parameter direction is normalized by the standard deviation $\varepsilon_{p_i}$ of the reconstruction of each parameter $p_i$
\begin{equation}
    \label{eq:distance}
    d = \sqrt{ \sum_{i=1}^M \left(\frac{p_i - p_i^{\ast}}{\varepsilon_{p_i^{\ast}}}\right)^2 }\,.
\end{equation}
The standard deviations $\varepsilon_{p_i^{\ast}}$ are determined from the Jacobian matrix of $\vec{f}(\vec{p})$ for $\vec{p}=\vec{p}^{\ast}$~\cite{Henn:12}. 
As a convergence criterion, we consider the case that a distance $d \leq \num{e-1}$ has been reached, i.e. that $\vec{p}$ has converged to within \SI{10}{\percent} of the reconstruction uncertainty.

\subsection{Problem}

\label{sec:problem}

\begin{figure}
    \centering
    \begin{minipage}[]{0.55\textwidth}
    \begin{tabular}{ccc}
    \toprule
    Parameter & Certified value & Parameter range \\
    \midrule
    $\beta_{1}$ & \num{699.64 \pm 16.30} & $[ \num{100}, \num{1000}]$ \\
    $\beta_{2}$ & \num{5.28 \pm 2.08} & $[\num{1}, \num{10}]$ \\
    $\beta_{3}$ & \num{0.76 \pm 0.20} & $[\num{0.1}, \num{1}]$ \\
    $\beta_{4}$ & \num{1.28 \pm 0.69} & $[\num{1}, \num{10}]$ \\
    \bottomrule
    \end{tabular}    
    \end{minipage}
    \begin{minipage}[]{0.4\textwidth}
\begin{tikzpicture}

\begin{axis}[
font=\small,
height=0.2 \textheight,
legend cell align={left},
legend style={
  fill opacity=0.8,
  draw opacity=1,
  text opacity=1,
  at={(0.03,0.97)},
  anchor=north west,
  draw=white!80!black
},
legend style={fill opacity=0.8, draw opacity=1, text opacity=1, at={(0.97,0.03)}, anchor=south east, draw=white!80!black},,
tick align=outside,
tick pos=left,
width=\columnwidth,
x grid style={white!69.0196078431373!black},
xlabel={Growth time (a.u.)},
xmin=0.3, xmax=15.7,
xtick style={color=black},
y grid style={white!69.0196078431373!black},
ylabel={Onion bulb wt. (a.u.)},
ymin=-19.3625, ymax=760.3725,
ytick style={color=black}
]
\addplot [only marks, line width=0pt, black, mark=+, mark size=3, mark options={solid}]
table {%
1 16.0799999237061
2 33.8300018310547
3 65.8000030517578
4 97.1999969482422
5 191.550003051758
6 326.200012207031
7 386.869995117188
8 520.530029296875
9 590.030029296875
10 651.919982910156
11 724.929992675781
12 699.559997558594
13 689.960021972656
14 637.559997558594
15 717.409973144531
};
\addlegendentry{Data points}
\addplot [semithick, black]
table {%
1 20.3018836975098
1.13999998569489 22.0409107208252
1.27999997138977 23.9263916015625
1.41999995708466 25.9701442718506
1.55999994277954 28.1848335266113
1.70000004768372 30.5840167999268
1.8400000333786 33.1821670532227
1.98000001907349 35.9947128295898
2.11999988555908 39.0380706787109
2.25999999046326 42.3296394348145
2.40000009536743 45.887825012207
2.53999996185303 49.7320365905762
2.6800000667572 53.8826484680176
2.8199999332428 58.3609580993652
2.96000003814697 63.1891288757324
3.09999990463257 68.3900985717773
3.24000000953674 73.987419128418
3.38000011444092 80.0051345825195
3.51999998092651 86.4675140380859
3.66000008583069 93.3988571166992
3.79999995231628 100.823112487793
3.94000005722046 108.763549804688
4.07999992370605 117.242317199707
4.21999979019165 126.279914855957
4.3600001335144 135.894683837891
4.5 146.102142333984
4.6399998664856 156.914337158203
4.78000020980835 168.339141845703
4.92000007629395 180.379516601562
5.05999994277954 193.032730102539
5.19999980926514 206.289764404297
5.34000015258789 220.134521484375
5.48000001907349 234.543426513672
5.61999988555908 249.484848022461
5.76000022888184 264.919006347656
6.03999996185303 297.065155029297
6.32000017166138 330.504150390625
7.15999984741211 432.430328369141
7.44000005722046 464.742828369141
7.57999992370605 480.245300292969
7.71999979019165 495.230529785156
7.8600001335144 509.648010253906
8 523.456237792969
8.14000034332275 536.623046875
8.27999973297119 549.125305175781
8.42000007629395 560.94873046875
8.5600004196167 572.08740234375
8.69999980926514 582.542907714844
8.84000015258789 592.323425292969
8.97999954223633 601.443115234375
9.11999988555908 609.9208984375
9.26000022888184 617.77978515625
9.39999961853027 625.045837402344
9.53999996185303 631.747436523438
9.68000030517578 637.914489746094
9.81999969482422 643.577819824219
9.96000003814697 648.768676757812
10.1000003814697 653.518127441406
10.2399997711182 657.856567382812
10.3800001144409 661.813842773438
10.5200004577637 665.418518066406
10.6599998474121 668.697937011719
10.8000001907349 671.678161621094
10.9399995803833 674.383666992188
11.0799999237061 676.837524414062
11.2200002670288 679.061279296875
11.3599996566772 681.074951171875
11.5 682.897155761719
11.6400003433228 684.545043945312
11.7799997329712 686.034362792969
11.9200000762939 687.379821777344
12.0600004196167 688.594665527344
12.1999998092651 689.691101074219
12.3400001525879 690.680358886719
12.4799995422363 691.572570800781
12.6199998855591 692.377014160156
12.7600002288818 693.102111816406
12.8999996185303 693.755554199219
13.039999961853 694.34423828125
13.1800003051758 694.87451171875
13.3199996948242 695.352111816406
13.460000038147 695.782165527344
13.6000003814697 696.169311523438
13.7399997711182 696.517822265625
13.8800001144409 696.83154296875
14.0200004577637 697.113830566406
14.1599998474121 697.367919921875
14.3000001907349 697.596557617188
14.4399995803833 697.80224609375
14.5799999237061 697.9873046875
14.7200002670288 698.15380859375
14.8599996566772 698.303527832031
15 698.438293457031
};
\addlegendentry{Cerfified fit result}
\end{axis}

\end{tikzpicture}
    \end{minipage}
    \caption{Left: certified values for the Rat43 dataset provided by the NIST Statistical Reference Database\cite{NIST_RAT43}, as well as parameter ranges chosen in the fitting process. Right: data points for the dataset and a certified reconstruction result, with a residual sum of squares $\chi^2_{\mathrm{min.}} = \num{8786.40 \pm 28.26}$. The parameters employed to create the certified fit result are found in the table on the left as certified values.}
    \label{fig:cert_vals}
\end{figure}
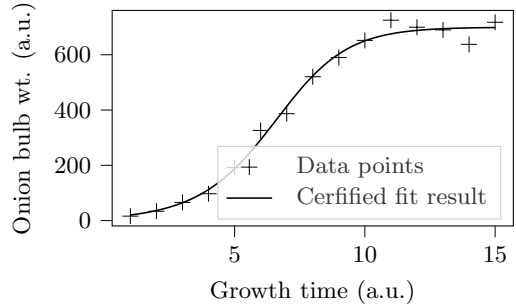

As benchmark problem we choose the Rat43 dataset, which is available in the NIST Standard Reference Database \cite{ratkowsky1983nonlinear,NIST_RAT43}. The dataset contains 15 discrete datapoints which describe the dry weight of onion bulbs and tops as a function of the growing time, as can be seen in \cref{fig:cert_vals}. The fit model is sigmoidal in nature and is controlled by four parameters which are gathered in the parameter vector $\boldsymbol{\beta}$, 
\begin{equation*}
    \label{eq:rat43_model}
    y = f(x, \boldsymbol{\beta}) = \frac{\beta_{1}}{1 + \exp{(\beta_{2} - \beta_{3}x)}^{\nicefrac{1}{\beta_{4}}}} \,.
\end{equation*}
The certified values given in the NIST repository are found in \cref{fig:cert_vals}. For the optimization a uniform prior with $\beta_{1} \in [100, 1000]$, $\beta_2 \in [1, 10]$, $\beta_3 \in [0.1, 1]$ and $\beta_4 \in [1, 10]$ was chosen. The number of datapoints and free parameters is such that this problem can pose as a stand-in for e.g. a small wavelength or angular scan, where a limited number of geometrical and experimental parameters is varied, such as for example in ellipsometry \cite{afraites2009application}. Without loss of generality we assume unit variance for each data channel, that is $\eta^{2}_{i} = 1 \, \forall i$. The regressor $x$ is varied between $\num{1}$ and $\num{15}$, yielding a vector $\vec{y}$ of length $\num{15}$.

Advantages of this analytical benchmark approach are the availability of analytical derivatives, the lack of discretization errors of e.g. finite element methods and a very short and controlled actual calculation time.

Due to their computational overhead, BO methods are best employed when used on expensive objective functions. To emulate an expensive model in the chosen analytical problem, a sleep call of ten seconds is inserted into the function evaluation.

\subsection{Reconstruction results}
\label{sec:results}

\begin{figure}
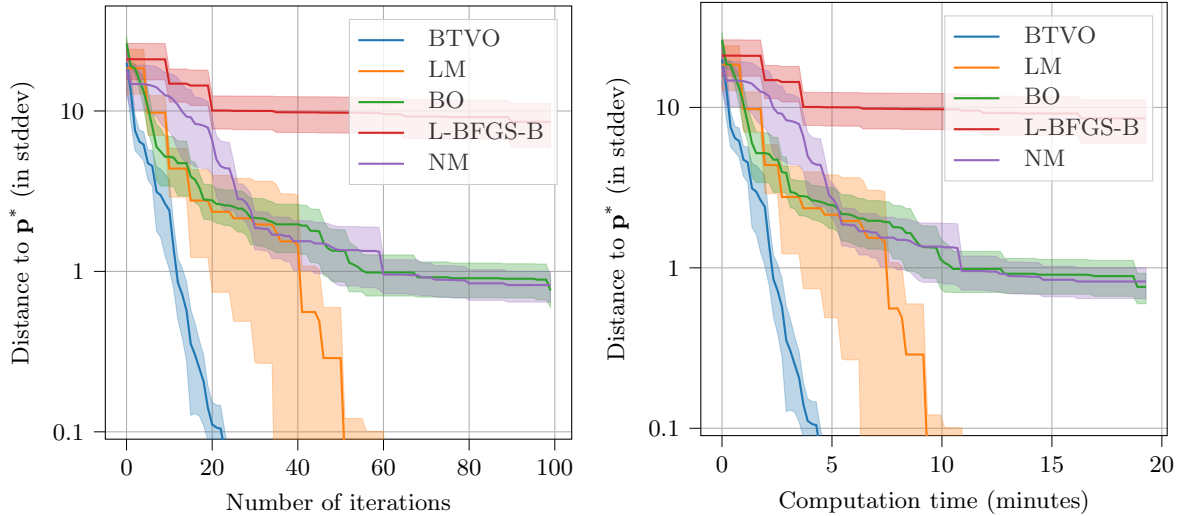

    \centering
    \begin{subfigure}{0.45\textwidth}
        \input{figures/rat43/tikz_no_derivatives_iterations_distance_mean_std}
    \end{subfigure}
    \begin{subfigure}{0.45\textwidth}
        \input{figures/rat43/tikz_no_derivatives_minutes_distance_mean_std}
    \end{subfigure}
    \caption{Shown are the results of the different optimization benchmarks: the average reconstruction result as a solid line and its standard deviation as shaded are around the average. The optimization metric is the distance of the reconstructed parameter $\vec{p}$ from the optimal parameter $\vec{p}^{\ast}$ in units of the standard deviations of the optimally reconstructed parameter for the BTVO with derivatives, as defined in \cref{eq:distance}. The proposed BTVO method outperforms every other optimization scheme, despite the computational overhead.}
    \label{fig:rat43_benchmark_no_derivatives_iterations_distance}
\end{figure}

The results of the different optimization benchmarks are shown in \cref{fig:rat43_benchmark_no_derivatives_iterations_distance}. We can clearly separate the performance of the optimization schemes into those which are capable of native least squares regression, and those which require a scalarization of the objective function. The native least squares methods both manage to achieve reconstruction results of \SI{10}{\percent} and less, well within the reserved optimization budget. Neither the L-BFGS-B, the NM, nor the conventional BO, are able to reconstruct the model parameters with the same accuracy in the optimization budget.

L-BFGS-B and LM both create a local approximation of the curvature of the objective function. L-BFGS-B uses an averaged approximation of the Hessian of the scalarized objective function, while LM creates a local approximation of the Jacobian of all of the outputs of the objective function. To achieve this, both methods rely on a finite differences scheme. A telltale sign of this is the steps that can be seen in \cref{fig:rat43_benchmark_no_derivatives_iterations_distance} and \cref{fig:rat43_benchmark_derivatives_iterations_distance} for the two methods.
Since L-BFGS-B determines the gradient of the objective function with a finite difference approximation and builds up an aproxiamtion of the Hessian over several iterations, it relies on more inaccurate approximations of the curvature in comparison to the LM approach. This explains the poorer performance of L-BFGS-B for performing a least square regression.

The NM method does not attempt to create an approximation of the local curvature of the objective function. Instead, it places a $M+1$ dimensional simplex in the parameter space, where $M$ is dimensionality of $\vec{p}$. The vertices of this simplex are used to create a model of the objective function. At each iteration of the scheme one vertex positions is updated with the goal of converging into a (local) minimizer of the objective. As no iterations are spent for determining finite differences, NM performs better than L-BFGS-B.

The BO method constructs a surrogate model of the scalarized objective function as detailed in \cref{sec:bo}. At the beginning of the optimization it outperforms NM. After about \num{30} iterations however, this advantage is lost, and BO and NM perform equally well. Of these three methods, only BO and NM achieve a reconstruction result within one standard deviation, while L-BFGS-B barely manages to reconstruct the parameters within \num{10} standard deviations. However, as they operate on the scalarized version of the objective function, neither L-BFGS-B, NM, nor BO, are apt methods to solve a least square regression problem.

The two native least squares methods, BTVO and LM, perform much better than the other reconstruction algorithms. They are capable of reconstructing the fit parameters to less than \SI{10}{\percent} of the chosen metric. Because LM has to first construct a Jacobian by means of a finite difference scheme, it is at a clear disadvantage over the BTVO. The surrogate model that is constructed during the optimization process with the BTVO containts these higher order terms implicitly. This speeds up the optimization considerably. As such, the BTVO reaches the desired reconstruction result within \num{24} iterations (corresponding to approximately \SI{4.5}{\minute}), while the LM algorithm requires \num{52} iterations ($\approx \SI{9.3}{\minute}$).

A list of reconstructed parameters is given in \cref{tab:reconstruction_results}. With exception of L-BFGS-B, all reconstructed parameters are well within the given standard deviation given as certified results by the NIST~\cite{NIST_RAT43}.

\subsubsection{Including derivative information}
\label{sec:results_derivatives}

\begin{figure}
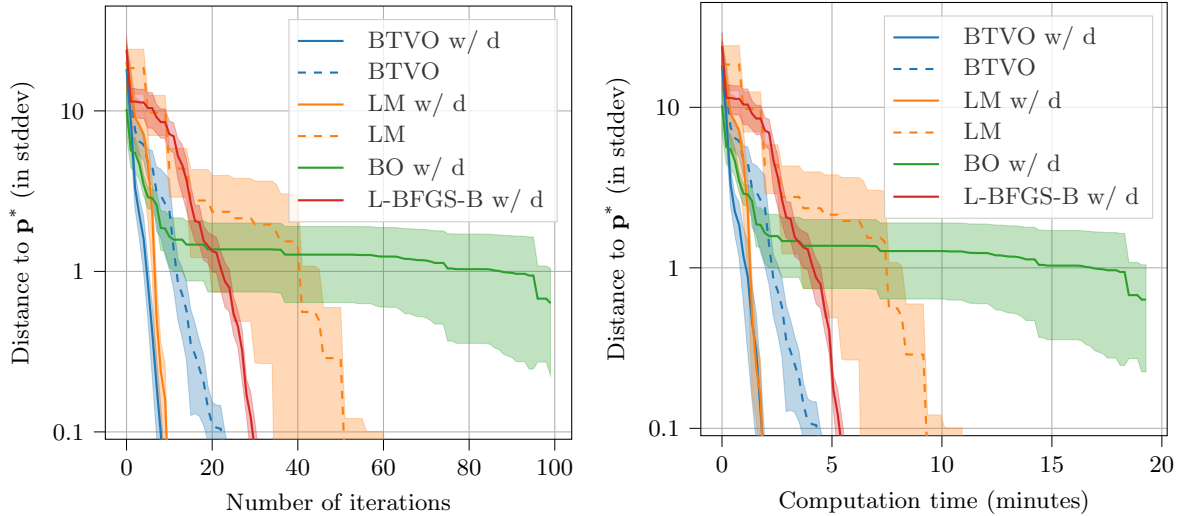

    \centering
    \begin{subfigure}{0.45\textwidth}
        \input{figures/rat43/tikz_derivatives_iterations_distance_mean_std}
    \end{subfigure}
    \begin{subfigure}{0.45\textwidth}
        \input{figures/rat43/tikz_derivatives_minutes_distance_mean_std}
    \end{subfigure}
    \caption{Shown are the results of the different optimization benchmarks or the case that derivative information is available: the average reconstruction result as a solid line and its standard deviation as shaded are around the average. The optimization metric is the distance of the reconstructed parameter $\vec{p}$ from the optimal parameter $\vec{p}^{\ast}$ in units of the standard deviations of the optimally reconstructed parameter for the BTVO with derivatives, as defined in \cref{eq:distance}. If derivatives are available, only the BO method is not able to reconstruct the model parameters to the desired accuracy.}
    \label{fig:rat43_benchmark_derivatives_iterations_distance}
\end{figure}

Since we are benchmarking an analytical objective function, derivatives with respect the all parameters are available. This can be used to speed up the parameter reconstruction considerably. Note that in some instances, this information may not be accessible, or only for a limited number of model parameters. Therefore the following results represent an idealized situation that is not necessarily found when trying to perform a least squares fit. 

In \cref{fig:rat43_benchmark_derivatives_iterations_distance} we  show the results of the optimization benchmarks for the methods that can directly utilize derivative information. For easy comparison to the derivative-free methods, BTVO and LM without derivatives are listed as well.

Since we provide the derivatives with respect to all parameters directly, LM and L-BFGS-B do not need to calculate a local approximation to the curvature by means of finite differences. Both schemes greatly benefit from the improved accuracy, but especially L-BFGS-B now also manages to reconstruct the model parameters to the desired accuracy, and well within the optimization budget. Also the convergence of BO improves with the use of derivatives. However, the convergence rate degrades after parameters within one standard deviation to the minimum ($d = 1$) have been found. We attribute this behaviour to the fact that the expected improvement becomes very small for $d \lessapprox 1$ and thus other parts of the parameter space are probed within the BO framework.

The substantial difference between the reconstruction performance of BTVO and LM has vanished with the availability of direct derivative information. Looking at the iterations that were required, BTVO took \num{9} and LM \num{11} calls of the objective function to achieve the desired accuracy. This slight headstart for the BTVO disappears when we consider the total computation time. Here, both methods achieved the desired accuracy in just under \SI{2}{\minute}. This is of course due to the computational overhead of the BTVO method.

Comparing these methods to their derivative-free counterparts, the BTVO with derivatives has achieved the desired accuracy in approximately a third of the iterations that it took without derivatives, and LM has managed the same thing in about a fifth of the iterations.

A list of reconstructed parameters is presented in \cref{tab:reconstruction_results}. All reconstructed parameters are well within the given standard deviation given as certified results by the NIST~\cite{NIST_RAT43}.

\begin{table}[]
    \centering
    \begin{tabular}{crcccccc}
    \toprule
    \multicolumn{2}{r}{Optimization method} & $\beta_1$ & $\beta_2$ & $\beta_3$ & $\beta_4$ & \multicolumn{2}{c}{Mean Its./Time} \\
    \midrule
    \multirow{5}{*}{(Without derivatives)} 
    & BTVO & $\num{ 699.6 }$ & $\num{ 5.3 }$ & $\num{ 0.8 }$ & $\num{ 1.3 }$ & 24 & \SI{4.5}{\minute} \\
    & LM & $\num{ 699.6 }$ & $\num{ 5.3 }$ & $\num{ 0.8 }$ & $\num{ 1.3 }$ & 52 & \SI{9.3}{\minute} \\
    & BO & $\num{ 702.7 }$ & $\num{ 4.9 }$ & $\num{ 0.7 }$ & $\num{ 1.1 }$ & \multicolumn{2}{c}{$\varnothing$} \\
    & L-BFGS-B & $\num{ 668.9 }$ & $\num{ 7.9 }$ & $\num{ 1.0 }$ & $\num{ 2.3 }$ & \multicolumn{2}{c}{$\varnothing$} \\
    & NM & $\num{ 697.5 }$ & $\num{ 5.8 }$ & $\num{ 0.8 }$ & $\num{ 1.4 }$ & \multicolumn{2}{c}{$\varnothing$} \\
    \midrule
    \multirow{4}{*}{(With derivatives)}
    & BTVO & $\num{ 699.6 }$ & $\num{ 5.3 }$ & $\num{ 0.8 }$ & $\num{ 1.3 }$ & 9 & \SI{1.9}{\minute} \\
    & LM & $\num{ 699.6 }$ & $\num{ 5.3 }$ & $\num{ 0.8 }$ & $\num{ 1.3 }$ & 11 & \SI{1.9}{\minute} \\
    & BO & $\num{ 699.4 }$ & $\num{ 5.3 }$ & $\num{ 0.8 }$ & $\num{ 1.3 }$ & \multicolumn{2}{c}{$\varnothing$} \\
    & L-BFGS-B & $\num{ 699.6 }$ & $\num{ 5.3 }$ & $\num{ 0.8 }$ & $\num{ 1.3 }$ & 31 & \SI{5.5}{\minute} \\
    \midrule
    & Certified results & $\num{ 699.6 \pm 16.3 }$ & $\num{ 5.3 \pm 2.1 }$ & $\num{ 0.8 \pm 0.2 }$ & $\num{ 1.3 \pm 0.7 }$ & & \\
    \bottomrule
    \end{tabular}
    \caption{The reconstructed parameters sorted by reconstruction scheme used. Also shown is the reconstruction performance in terms of the mean number of calls to the objective function (Its.) and mean time that was required (Time) for the reconstruction to achieve the desired accuracy. A discussion and explanations for deviations from the certified values can be found in \cref{sec:results} and \cref{sec:results_derivatives}.}
    \label{tab:reconstruction_results}
\end{table}

\section{Summary and outlook}
\label{sec:outlook}

We have shown that the proposed extension to BO can be successfully employed in the reconstruction of model parameters of an analytical non-linear least-square problem. For a specific test problem, it performs almost twice as good as an established least-square algorithm when no derivatives are available, and at least similarly well if derivatives are at hand. We have also shown that it outperforms the conventional BO, which has been successfully employed in the reconstruction of parameters in optical metrology~\cite{andrle2019grazing,schneider2019benchmarking}. We are therefore preparing a publication in which we are investigating the applicability of the BTVO to a comparable problem from optical metrology.

Furthermore, we expect that the BTVO method has important advantages for the error estimation of the reconstructed parameters. BTVO is based on a non-linear surrogate model for each of the outputs of the objective function. This surrogate model can be evaluated quickly to predict the fitting error $\chi^2(\vec{p})$ close to best parameter set $\vec{p}^{\ast}$ with good accuracy. The surrogate can thus be exploited to perform a Markow Chain Monte Carlo (MCMC) sampling of the posterior density~\cite{klauenberg2016markov,farchmin2020efficient}. Performing such a sampling of the actual objective function is computationally very
expensive and can take many days~\cite{rasmussen2015novel}. The use of an accurate surrogate model has thus the potential to accelerate MCMC analyses considerably. 
We plan to extend our current work in this direction.

Further, the proposed BTVO method does not take covariances between the different data channels into account. This could potentially further enhance the reconstruction ability. 
A possible approach, introduced by Matsui et al., includes the covariance structure into the GPs~\cite{matsui2019bayesian}. 
We assume that a corresponding approach 
could be applied to the proposed BTVO method.

\acknowledgments 
 
This project is funded by the German Federal Ministry of Education and Research (BMBF, project number 05M20ZAA, siMLopt, and project number 01IS20080A, SiM4diM). 
We further acknowledge funding from the German Federal Ministry of Education and Research (BMBF Forschungscampus  MODAL, project number 05M20ZBM) and from the German Federal Ministry for Economic Affairs and Energy (BMWi, project number 50WM2067, Optimal-QT).

\bibliography{bibl} 
\bibliographystyle{spiebib} 

\end{document}